\documentclass[aps,prc,twocolumn,reprint,floatfix,superscriptaddress,showpacs]{revtex4-1}
\usepackage{graphicx}
\usepackage[pdftex,colorlinks,citecolor=blue,bookmarks]{hyperref}
\usepackage{bm}
\usepackage{longtable}
\begin{document}
\title{Calculation of nuclear matrix elements in neutrinoless double electron capture}

\author{Tom\'as R. Rodr\'iguez} 
\affiliation{Technische Universit\"at Darmstadt, D-64289 Darmstadt, Germany} 
\affiliation{GSI Helmholtzzentrum
  f\"ur Schwerionenforschung, D-64291 Darmstadt, Germany} 
\author{Gabriel Mart\'inez-Pinedo} 
\affiliation{Technische Universit\"at Darmstadt, D-64289 Darmstadt, Germany} 
\affiliation{GSI Helmholtzzentrum
  f\"ur Schwerionenforschung, D-64291 Darmstadt, Germany}
\date{\today} 
\begin{abstract}
We compute nuclear matrix elements for neutrinoless double electron capture on $^{152}$Gd, $^{164}$Er and $^{180}$W nuclei. Recent precise mass measurements for these nuclei have shown a large resonance enhancement factor that makes them the most promising candidates for observing this decay mode. We use an advanced energy density functional method which includes beyond mean-field effects such as symmetry restoration and shape mixing. Our calculations reproduce experimental charge radii and $B(E2)$ values predicting a large deformation for all these nuclei. This fact reduces significantly the values of the NMEs leading to half-lives larger than $10^{29}$ years for the three candidates. 
\end{abstract}
\maketitle
\section{Introduction}
\begin{figure}[b!]
\begin{center}
  \includegraphics[width=\columnwidth]{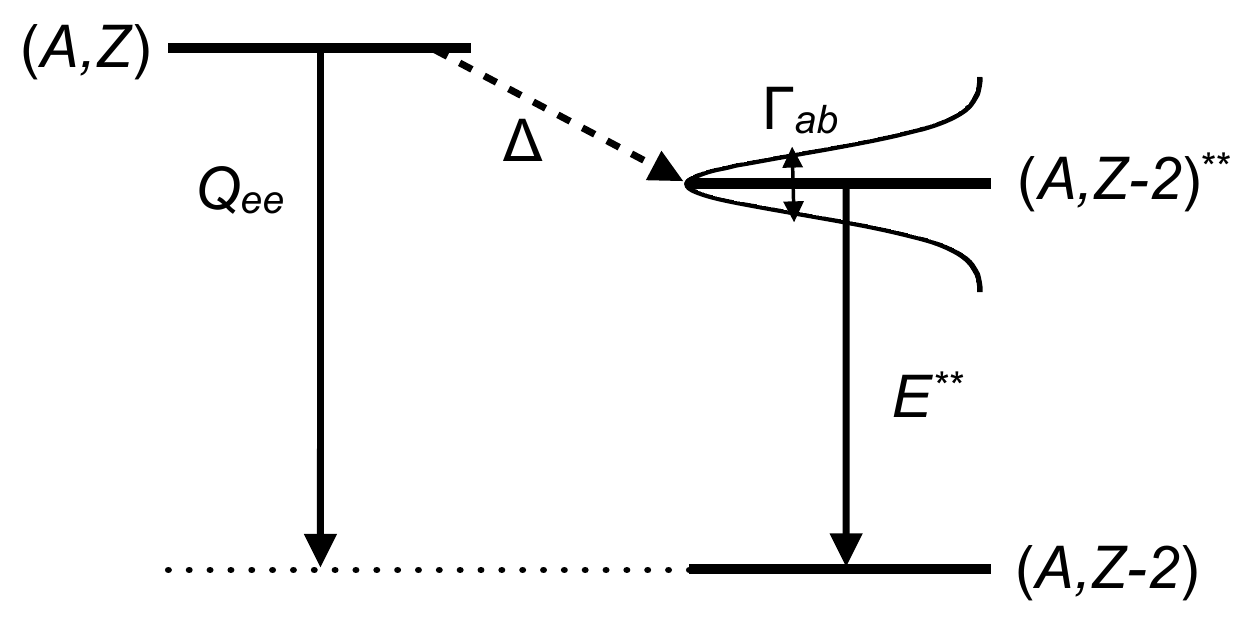}
\end{center}
\caption{Energetics of the $0\nu$ECEC process. See text for details.}\label{Fig1}
\end{figure}

The nature of the neutrino as a Majorana or Dirac particle is still an open question. Experimental evidence of lepton-number violation weak processes such as neutrinoless double beta decay ($0\nu\beta\beta$) or double electron capture ($0\nu$ECEC) would determine unambiguously the Majorana character of this elementary particle. In particular, in $0\nu$ECEC a Majorana neutrino is exchanged in the capture of two electrons ($e^{-}$) from the inner shells of an atom with mass number $A$ and number of protons  $Z$: 
\begin{equation}
e^{-}+e^{-}+(A,Z)\rightarrow (A,Z-2)^{**}
\end{equation}
The daughter atom $(A,Z-2)^{**}$ is unstable and decays to its ground state $(A,Z-2)$ by X-rays and/or Auger electrons emission from the electronic excitation and by $\gamma$-ray emission if the nucleus is in an excited state. In Fig.~\ref{Fig1} we show a schematic view of the process and its energetics. The capture rate $\lambda_{0\nu ee}$ can be written as~\cite{NPB_223_15_1983}:
\begin{equation}
\lambda_{0\nu ee}= \frac{|V_{ab}|^{2}}{\Delta^{2}+\frac{1}{4}\Gamma_{ab}^{2}}\Gamma_{ab}
\label{rate}
\end{equation}
where $\Delta\equiv Q_{ee}-E^{**}$, $Q_{ee}$ is the difference between the initial and final atomic masses, $E^{**}$ and $\Gamma_{ab}$ are the energy and the width of the excited state of the daughter atom with two electronic holes in states $a,b$ and an excited nucleus, and $V_{ab}$ is the transition amplitude between the initial and final states~\cite{NPA_859_140_2011}:
\begin{equation}
V_{ab}=m_{\beta\beta}(G_{F}\cos\theta_{C})^{2}\frac{g^{2}_{A}}{4\pi R}\langle F_{ab}\rangle M^{0\nu}
\label{transition_ME}
\end{equation}
with $m_{\beta\beta}$, $G_{F}$, $\theta_{C}$, $g_{A}=1.25$, $R$ being the effective Majorana neutrino mass, the Fermi constant, the Cabbibo angle, the axial-vector coupling constant and the nuclear radius respectively; $\langle F_{ab}\rangle$ is the lepton part of the matrix element that takes into account the atomic electron wave functions of the states $a,b$ and $M^{0\nu}$ is the nuclear matrix element (NME)~\cite{NPA_859_140_2011}.
Equation~\ref{rate} shows a resonant behavior with an increase of the capture rate whenever $\Delta\sim\Gamma$ is fulfilled, otherwise the process is very much suppressed. This is a very restricted condition to search for isotopes with a potential $0\nu$ECEC because typical values for the widths of the atomic states are of the order of electron volts (eV) while $Q_{ee}$ is found by subtracting two quantities which are of the order of MeV. 
Recent experiments using Penning traps have reported extremely precise measurements of $Q_{ee}$ values for the most promising $0\nu$ECEC candidates where the initial and final nuclei are in their ground states, namely, $^{152}$Gd~\cite{PRL_106_052504_2011}, $^{164}$Er~\cite{PRL_107_152501_2011} and $^{180}$W~\cite{NPA_875_1_2012}. These experiments have shown rather large resonance enhancement factors that could lead to $0\nu$ECEC half-lives between $10^{26-30}$~years$/(m_{\beta\beta}[eV])^{2}$~\cite{NPA_875_1_2012}. However, this half-lives are based on NMEs computed using spherical quasiparticle random phase approximation (QRPA) for mother and daughter nuclei~\cite{NPA_859_140_2011}. Recently Fang et \textit{al.}~\cite{arXiv:1111.6862v1} have extended these calculations to deformed shapes and found a substantial suppression of the NMEs that leads to longer half-lives. However, in these calculations the deformation is a free parameter that is not consistently determined within the model.

In this work we use an energy density functional (EDF) method including beyond-mean field correlations to evaluate the $0\nu$ECEC NMEs for these isotopes. This method was originally developed to study in a self-consistent manner the role of deformation, shape mixing and pairing correlations in $0\nu\beta\beta$ NMEs~\cite{PRL_105_252503_2010,PPNP_66_436_2011}. Here, a density-dependent effective two body interaction between the constituent nucleons is assumed as the starting point. Then, single particle energies, residual interactions beyond the mean field, pairing and deformation content of the nuclei, transitions and decays rates, etc., are deduced self-consistently from solving the associated quantum many-body problem. In addition, mixing of different shapes is also allowed within this formalism. This is a different perspective that the one used in QRPA calculations~\cite{arXiv:1111.6862v1}, where the several free parameters of the model are adjusted specifically for each capture. In particular, the deformation of initial and final nuclei are fitted to reproduce either the spectroscopic quadrupole moment of the $2^{+}_{1}$ state or from the reduced transition probabilities $B(E2)$. The effects of shape mixing, particle number and angular momentum restoration are not considered. In addition, pairing gaps are also found by fitting at the BCS level experimental even-odd mass differences. Finally, single particle energies and residual interactions do not come from the same underlying interaction, and an additional renormalization is required adding two more adjustable parameters to the model. 
\section{Theoretical framework}
We now summarize the formalism that we use to compute the NMEs considering that the initial and final nuclei are in their ground states. First of all, $M^{0\nu}$ can be expressed as the sum of the so-called Fermi (F), Gamow-Teller (GT) and tensor (T) parts~\cite{RMP_80_481_2008}:
\begin{equation}
M^{0\nu}=-\left(\frac{g_{V}}{g_{A}}\right)^{2}M^{0\nu}_{F} +M^{0\nu}_{GT}-M^{0\nu}_{T}  
\label{MMM}
\end{equation}
with $g_{V}=1$. 
In this work we neglect the tensor term that it is estimated to be small~\cite{NPA_818_139_2009,PRC_75_051303_2007}. The other terms in Eq. \ref{MMM} are written as the overlaps of two-body operators between initial and final states, $M^{0\nu}_{F/GT}=\langle
0^{+}_{f}|\hat{M}^{0\nu}_{F/GT}|0^{+}_{i}\rangle$, with:
\begin{equation}
  \hat{M}^{0\nu}_{F}=\hat{V}_{F}\hat{t}^{(1)}_{+}\hat{t}^{(2)}_{+},\quad
  \hat{M}^{0\nu}_{GT}=\hat{V}_{GT} (\hat{\bm{\sigma}}^{(1)} \cdot
  \hat{\bm{\sigma}}^{(2)})\hat{t}^{(1)}_{+}\hat{t}^{(2)}_{+}   
  \label{NME}
\end{equation} 
Here $\hat{t}_{+}$ is the isospin ladder operator that changes protons into neutrons and $\hat{\bm{\sigma}}$ are the Pauli matrices acting on the spin part. The spatial part of the wave functions is affected by the so-called neutrino potentials $\hat{V}_{F/GT}$ that include nucleon finite size corrections~\cite{PRC_60_055502_1999}, radial short-range correlations within the unitary correlation operator Method (UCOM)~\cite{NPA_632_61_1998,PRC_77_045503_2008} and higher order currents~\cite{PRC_60_055502_1999}. Detailed expressions for these operators can be found in Refs.~\cite{PRC_60_055502_1999,NPA_818_139_2009}.

Initial ($i$) and final ($f$) states are found by using configuration mixing techniques based on the generator coordinate method (GCM) with particle number and angular momentum restoration of Hartree-Fock-Bogoliubov type (HFB) mean-field wave functions~\cite{RingSchuck,RMP_75_121_2003,NPA_709_467_2001,PRL_99_062501_2007}:
\begin{equation}
|I^{+}_{i/f}\rangle=\sum_{\beta_{2}}g^{I}_{i/f}(\beta_{2})|\Psi^{I}_{i/f}(\beta_{2})\rangle
\label{GCM_WF}
\end{equation}
where the axial quadrupole deformation $\beta_{2}$ is used as the generating coordinate and $g^{I}_{i/f}(\beta_{2})$ are the coefficients of the linear combination which are obtained by solving the corresponding Hill-Wheeler-Griffin (HWG) equations~\cite{RingSchuck}. The set of wave functions  $|\Psi^{I}_{i/f}(\beta_{2})\rangle$ are defined as:
\begin{equation}
|\Psi^{I}_{i/f}(\beta_{2})\rangle=P^{I}P^{N_{i/f}}P^{Z_{i/f}}|\Phi_{i/f}(\beta_{2})\rangle
\label{PNAMP_WF}
\end{equation} 
being $P^{I}$, $P^{N}$ and $P^{Z}$ the projection operators on to angular momentum $I$ and number of neutrons and protons, respectively; $|\Phi_{i/f}(\beta_{2})\rangle$ are HFB type mean-field wave functions that are found by using the variation after particle number projection method~\cite{NPA_696_467_2001}. We have used two parametrizations of the Gogny force as the underlying density-dependent interaction, namely, the most widely used in nuclear structure calculations D1S~\cite{NPA_428_23_1984} and the most recent one D1M~\cite{PRL_102_242501_2009} that is designed to give the best fit to experimental masses in the whole nuclear chart.
Experimental observables such as nuclear masses, radii, spectra, transition rates, etc. can be directly compared to the expectation values calculated from the GCM wave functions given in Eq.~\ref{GCM_WF}. In addition, the overlaps  between the projected states defined in Eq.~\ref{PNAMP_WF} of the corresponding operators give the dependence on the quadrupole deformation of such observables. Finally, probability distributions for the GCM wave functions -Eq.~\ref{GCM_WF}- to have a given deformation, the so-called collective wave functions $F^{I}_{i/f}(\beta_{2})$, provide additional information about the intrinsic structure of these states and identify the relevant regions within the space parametrized by the deformation $\beta_{2}$ (see~\cite{PRL_105_252503_2010,PPNP_66_436_2011} and references therein for detailed expressions). In this work only HFB axial symmetric configurations are considered and these intrinsic wave functions are expanded in a single particle basis composed by eleven major spherical harmonic oscillator shells. 
\begin{table}[b!]
\begin{center}
\begin{tabular}{c|c|c|c|c|c}
\hline\hline
Isotope & & $BE/A$ & $\langle r^{2}_{c}\rangle^{1/2}$) & $E(2^{+})$ & $B(E2\downarrow)$ \\
 & & (MeV) & (fm) & (MeV) & (W.u) \\
\hline\hline

$^{152}$Gd & $\begin{array}{c} \mathrm{D1S} \\ \mathrm{D1M} \\ \mathrm{Expt} \end{array}$ & $\begin{array}{c} 8.212 \\ 8.181 \\ 8.233 \end{array}$ & $\begin{array}{c} 5.027 \\ 4.991 \\ 5.082 \end{array}$ & $\begin{array}{c} 0.227 \\ 0.228 \\ 0.344 \end{array}$ & $\begin{array}{c} 113 \\ 104 \\ 73 (7) \end{array}$ \\
\hline
$^{152}$Sm & $\begin{array}{c} \mathrm{D1S} \\ \mathrm{D1M} \\ \mathrm{Expt} \end{array}$ & $\begin{array}{c} 8.219 \\ 8.195 \\ 8.244 \end{array}$ & $\begin{array}{c} 5.061 \\ 5.024 \\ 5.084 \end{array}$ & $\begin{array}{c} 0.118 \\ 0.117 \\ 0.122 \end{array}$ & $\begin{array}{c} 204 \\ 186 \\ 144 (3) \end{array}$\\
\hline
$^{164}$Er & $\begin{array}{c} \mathrm{D1S} \\ \mathrm{D1M} \\ \mathrm{Expt} \end{array}$ & $\begin{array}{c} 8.118 \\ 8.088 \\ 8.149 \end{array}$ & $\begin{array}{c} 5.190 \\  5.154 \\ 5.238 \end{array}$ & $\begin{array}{c} 0.125 \\ 0.125 \\ 0.091\end{array}$ & $\begin{array}{c} 226 \\ 215 \\ 218 (7)\end{array}$ \\
\hline
$^{164}$Dy & $\begin{array}{c} \mathrm{D1S} \\ \mathrm{D1M} \\ \mathrm{Expt} \end{array}$ & $\begin{array}{c} 8.122 \\ 8.103 \\ 8.158 \end{array}$ & $\begin{array}{c} 5.181 \\ 5.149 \\ 5.221 \end{array}$ & $\begin{array}{c} 0.105 \\ 0.103 \\ 0.073 \end{array}$ & $\begin{array}{c} 231 \\ 225 \\ 209 (3)\end{array}$ \\
\hline
$^{180}$W & $\begin{array}{c} \mathrm{D1S} \\ \mathrm{D1M} \\ \mathrm{Expt} \end{array}$ & $\begin{array}{c} 7.982 \\ 7.955 \\ 8.025 \end{array}$ & $\begin{array}{c} 5.321 \\ 5.289 \\ 5.349 \end{array}$ & $\begin{array}{c} 0.137 \\ 0.136 \\ 0.093 \end{array}$ & $\begin{array}{c} 172 \\ 172 \\ $--$ \end{array}$ \\
\hline
$^{180}$Hf & $\begin{array}{c} \mathrm{D1S} \\ \mathrm{D1M} \\ \mathrm{Expt} \end{array}$ &  $\begin{array}{c} 7.987 \\ 7.968 \\ 8.034 \end{array}$ & $\begin{array}{c} 5.312 \\ 5.276 \\ 5.342 \end{array}$ & $\begin{array}{c} 0.133 \\ 0.130 \\ 0.104 \end{array}$ & $\begin{array}{c} 175 \\ 167 \\ 155 (5) \end{array}$ \\
\hline\hline
\end{tabular}
\end{center}
\caption{Theoretical and experimental data for ground state and spectroscopic properties of the nuclei involved in $0\nu$ECEC. First and second rows of each cell correspond to theoretical values calculated with Gogny D1S and D1M parametrizations respectively. Experimental data are taken from Refs.~\cite{ENSDF,NPA_729_337_2003,ADNDT_84_185_2004}}
\label{table1}
\end{table}
\section{Results}
We now present the results obtained for the $0\nu$ECEC processes $^{152}\mathrm{Gd}\rightarrow\mathrm{^{152}Sm}$, $^{164}\mathrm{Er}\rightarrow\mathrm{^{164}Dy}$ and $^{180}\mathrm{W}\rightarrow\mathrm{^{180}Hf}$. First, to check the reliability of the method applied to these nuclei, we compare in Table~\ref{table1} theoretical and experimental values for different observables such as: binding energy per particle and nuclear charge radius -ground state properties- or excitation energy of the first $2_{1}^{+}$ state and its electric quadrupole reduced transition probability $B(E2; 2^{+}_{1}\rightarrow0^{+}_{1})$ -spectroscopic properties. On the one hand, we see that the results provided by D1S and D1M parametrizations are almost identical, although D1S gives a bit higher binding energies, radii and $B(E2)$ values. On the other hand, theoretical predictions give a bit smaller binding energies and radii, larger $B(E2)$ values and, except for $A=152$, larger $2^{+}$ excitation energies than the experimental data. Nevertheless, the agreement with the experiments is very good, taking into account that the interactions are globally fitted and effective charges are not needed within this framework. Furthermore, the relative systematics for the pair of nuclei involved in the electron capture is reproduced, i.e., whenever a quantity is experimentally larger or smaller in one nucleus than in the other, the same trend is observed in the theoretical results. Other degrees of freedom not included in these calculations like, for example, triaxiality and time-reversal symmetry breaking could help to improve the agreement with the experimental data, especially for the excitation energies and transition probabilities. Work in this direction is in progress.
\begin{figure}[b]
\begin{center}
  \includegraphics[width=\columnwidth]{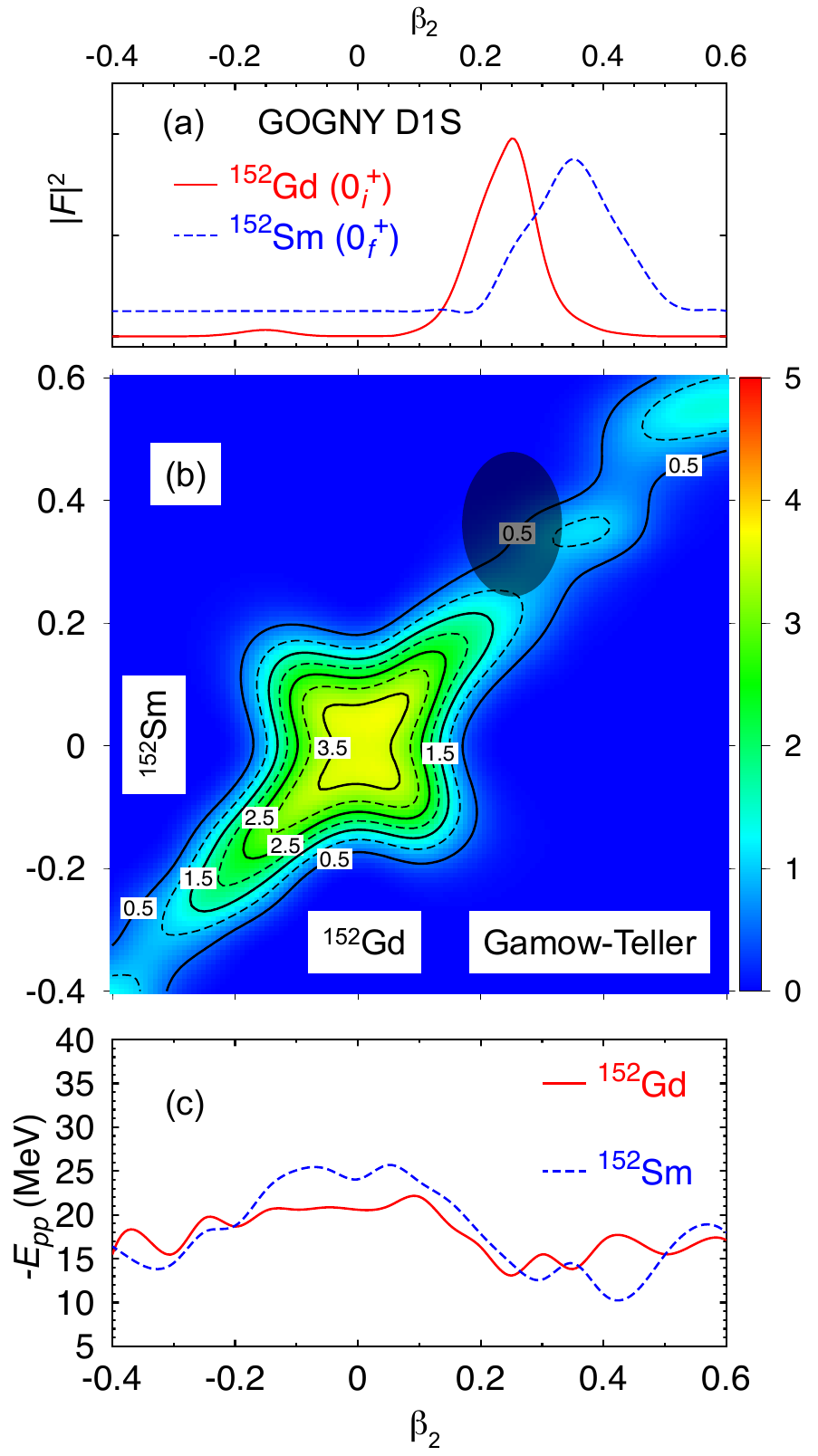}
\end{center}
\caption{(color online) (a) Collective wave functions ; (b) intensity of the Gamow-Teller part of the $0\nu$ECEC NME and (c) pairing energies as a function of the quadrupole deformation $\beta_{2}$ for $^{152}$Gd and $^{152}$Sm isotopes calculated with the Gogny D1S parametrization.}\label{Fig2}
\end{figure}
\begin{figure}[htb]
\begin{center}
  \includegraphics[width=\columnwidth]{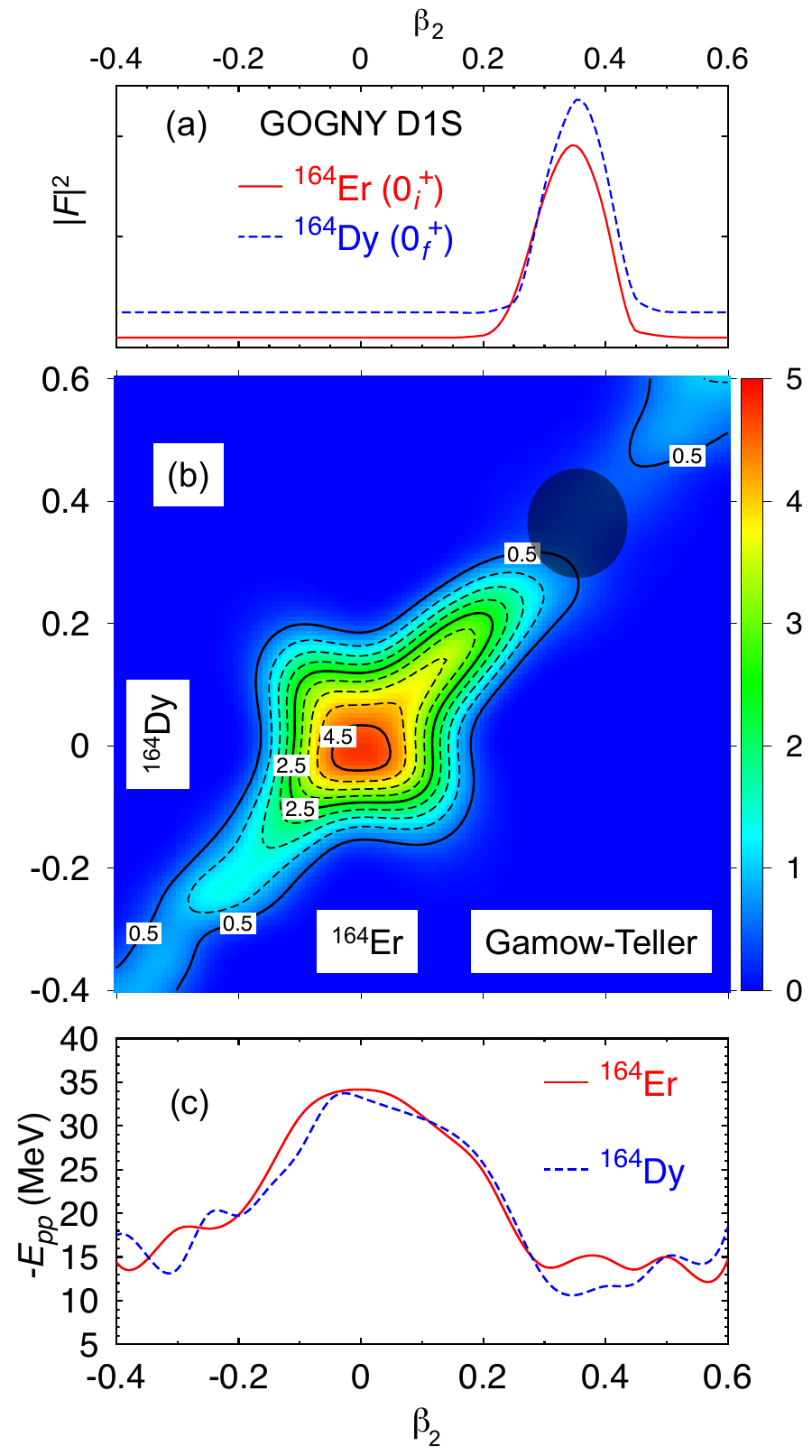}
\end{center}
\caption{(color online) Same as Fig.~\ref{Fig2} but for $^{164}$Er and $^{164}$Dy isotopes.}\label{Fig3}
\end{figure}
\begin{figure}[htb]
\begin{center}
  \includegraphics[width=\columnwidth]{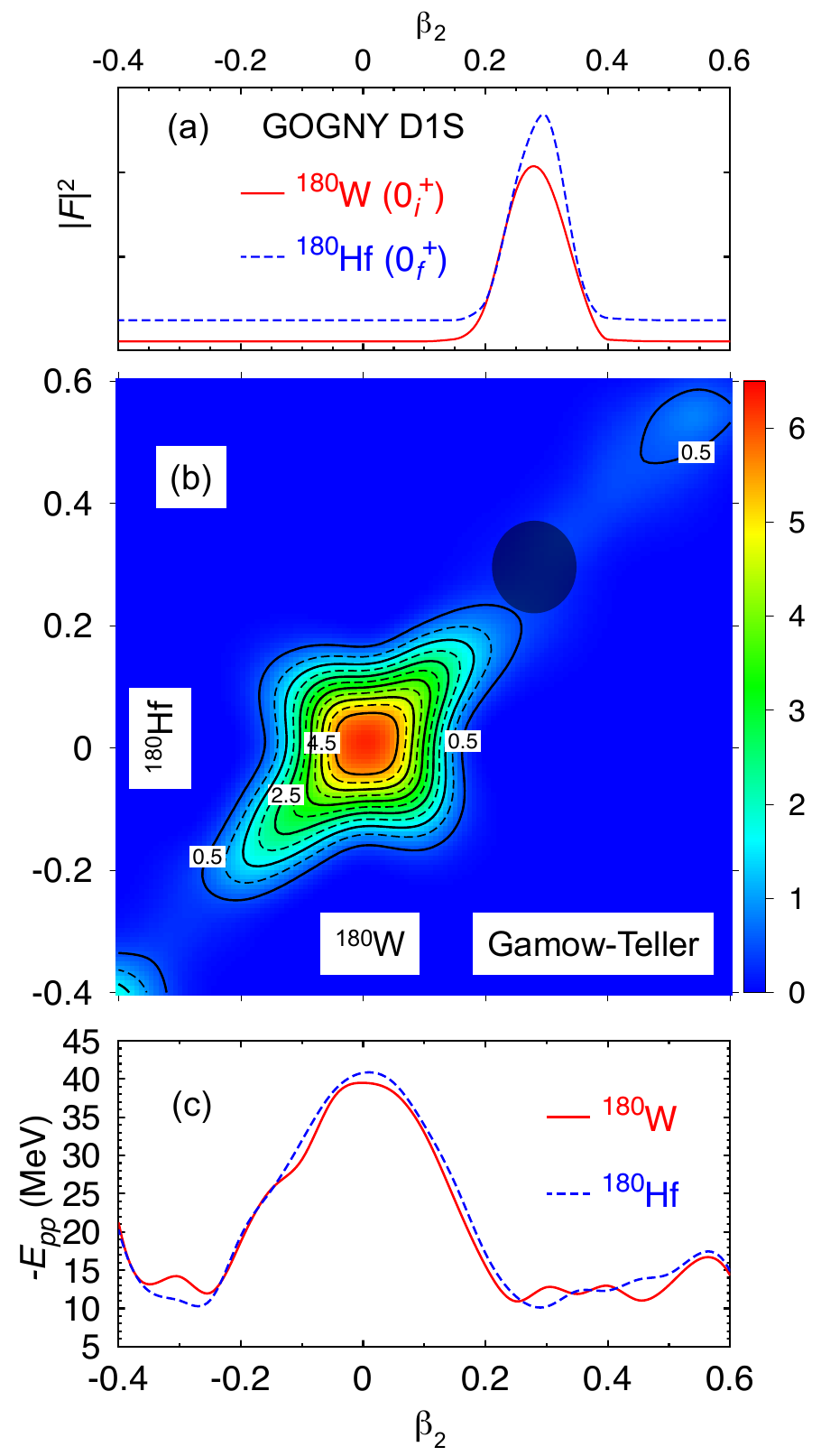}
\end{center}
\caption{(color online) Same as Fig.~\ref{Fig2} but for $^{180}$W and $^{180}$Hf isotopes.}\label{Fig4}
\end{figure}

All the isotopes studied in this work are open shell nuclei, both in protons and neutrons. This means that collective effects are expected to appear. In fact, most of these isotopes show experimentally rotational ground state bands~\cite{ENSDF} pointing out that quadrupole deformations play a significant role. The GCM method used here is particularly suitable to include this degree of freedom in a self-consistent manner. In Figs.~\ref{Fig2}(a)-~\ref{Fig3}(a)-~\ref{Fig4}(a) we plot the collective wave functions -distribution of probability of having the state a given deformation- as a function of the quadrupole deformation $\beta_{2}$ for the nuclei calculated in this work using the D1S parametrization (similar features are found for D1M). We observe that all the isotopes have a large prolate deformation. Smaller deformation and narrower distribution is obtained for $^{152}$Gd than for $^{152}$Sm. The maxima of the collective wave functions are found at $\beta_{2}\sim0.25$ and $\beta_{2}\sim0.35$ respectively. For the other two cases mother and daughter collective wave functions are very similar, having their maxima at $\beta_{2}\sim0.35$ for $A=164$ and $\beta_{2}\sim0.30$ for $A=180$. Therefore, the quadrupole degree of freedom has to be taken into account in the calculation of the $0\nu$ECEC rates. 

The dependence of the NMEs on the deformation can be obtained by calculating the overlaps of the operators defined in Eq.~\ref{NME} between the particle number and angular momentum projected states given in the expression~\ref{PNAMP_WF}~\cite{PRL_105_252503_2010}. We represent the intensity of the Gamow-Teller part of the NME as a function of the deformation of the mother and daughter nuclei in Figs.~\ref{Fig2}(b)-~\ref{Fig3}(b)-~\ref{Fig4}(b). The same qualitative behavior is found for the Fermi part and also for the D1M parametrization and they are not shown. The three cases display similar characteristics. Highest intensities are located around the spherical shape and along the diagonal $\beta^{\mathrm{ini}}_{2}=\beta^{\mathrm{fin}}_{2}$ between $\beta_{2}\in[-0.3,0.25]$ for $A=152,164$ and $\beta_{2}\in[-0.2,0.2]$ for $A=180$. The rest of configurations are very much suppressed. Therefore, $0\nu$ECEC is favored whenever the mother and daughter are nearly spherical in these cases. We also observe that the distribution of the NMEs as a function of the deformation is correlated to the pairing energy ($E_{pp}$) of the nuclei involved in the capture process (see Figs.~\ref{Fig2}(c)-~\ref{Fig3}(c)-~\ref{Fig4}(c)). Hence, whenever a maximum in the paring energy is present, also a maximum in the NME is obtained. The same correlation has been accounted for in the $0\nu\beta\beta$ NMEs~\cite{PRL_105_252503_2010,PPNP_66_436_2011} and using a seniority scheme~\cite{PRL_100_052503_2008,PRC_77_045503_2008}. 
However, although spherical shapes are favored by the $0\nu$ECEC operator, we see in Figs.~\ref{Fig2}(b)-~\ref{Fig3}(b)-~\ref{Fig4}(b) that the actual many-body wave functions explore only regions that corresponds to quite small intensities- see the shaded area. Therefore, the NMEs are strongly suppressed whenever the deformation of the nuclei is taken into account as it was already observed in the calculation of $0\nu\beta\beta$ NMEs of $^{150}$Nd~\cite{PRL_105_252503_2010}. A similar qualitative result is obtained in QRPA calculations~\cite{arXiv:1111.6862v1} although the suppression with respect to the spherical result is much larger in this work.
\begin{table}[t]
\begin{center}
\begin{tabular}{c|ccc|c}
\hline\hline
Nucleus & Param. & $M^{0\nu}$ & $T^{\mathrm{min}}_{1/2}(y)$  & $M^{0\nu}$ \\
 &  & EDF &  & QRPA~\cite{arXiv:1111.6862v1} \\
\hline\hline
$^{152}$Gd &  $\begin{array}{c} \mathrm{D1S} \\ \mathrm{D1M} \end{array}$ & $\begin{array}{c}1.07 \\ 0.89 \end{array}$ & $\begin{array}{c} 4.2\times10^{29} \\ 6.2\times10^{29} \end{array}$ & 3.23-2.67 \\ \hline
$^{164}$Er & $\begin{array}{c} \mathrm{D1S} \\ \mathrm{D1M} \end{array}$ &  $\begin{array}{c} 0.64 \\ 0.50 \end{array}$ & $\begin{array}{c} 1.3\times10^{34} \\ 2.1\times10^{34} \end{array}$ & 2.64-2.27 \\ \hline
$^{180}$W & $\begin{array}{c} \mathrm{D1S} \\ \mathrm{D1M} \end{array}$ & $\begin{array}{c} 0.58 \\ 0.38 \end{array}$ & $\begin{array}{c} 1.6\times10^{32} \\ 3.8\times10^{32} \end{array}$ & 2.05-1.79 \\ \hline
\hline
\end{tabular}
\end{center}
\caption{NMEs and estimated half-lives for most probable $0\nu ee$ captures.}
\label{table2}
\end{table}%

\begin{figure}[t]
\begin{center}
  \includegraphics[width=\columnwidth]{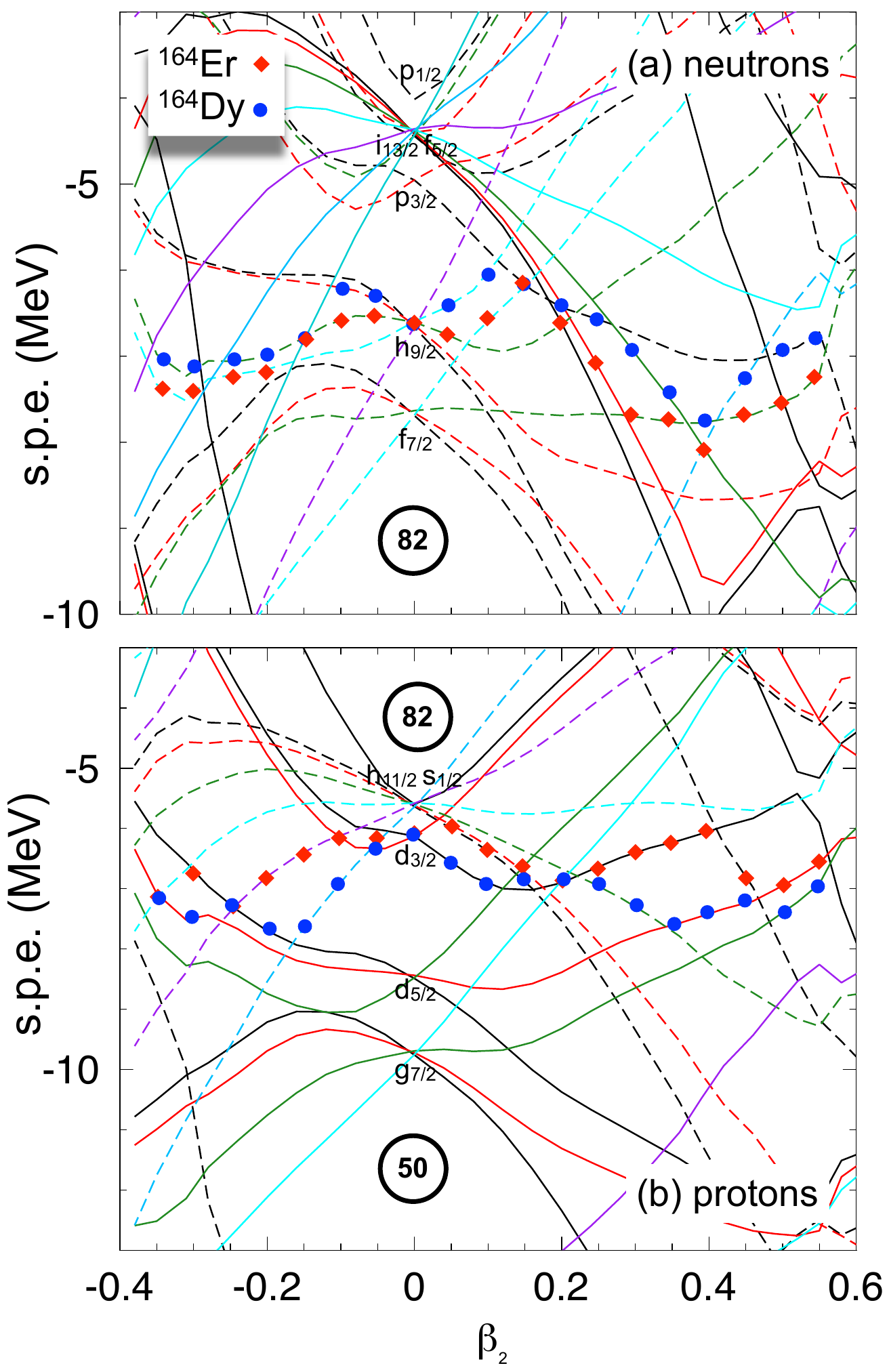}
\end{center}
\caption{(color online) Single particle energies (continuous and dashed lines represent positive and negative parity levels respectively) as a function of the deformation for $A=164$ for (a) neutrons and (b) protons. The red diamonds and the blue bullets are the Fermi levels for $^{164}_{\,\,\,68}$Er$_{96}$ and $^{164}_{\,\,\,66}$Dy$_{98}$ isotopes respectively.}\label{Fig5}
\end{figure}
To shed more light on the structure of the NMEs and pairing energies shown above, we plot in Fig.~\ref{Fig5} the single particle energies (s.p.e.) as a function of the deformation for the $A=164$ isotopes. A similar analysis applies also to the rest of isotopes studied here. In the spherical point ($\beta_{2}=0$) these levels have a degeneracy equal to $(2j+1)$. For $\beta_{2}\neq0$ this degeneracy is broken in $(j+1/2)$ doubly degenerated levels. In the following discussion we maintain the name of the spherical orbits for referring the different levels. It is well known that pairing energies are larger whenever the level density around the Fermi energy is larger. This is precisely what it is observed in the range of deformations $\beta_{2}\in[-0.2,0.2]$ where proton levels coming from the orbits $\pi d_{3/2}, \pi h_{11/2}, \pi s_{1/2}$ are very close in energy. For the neutrons, the fact that the spherical orbit $\nu h_{9/2}$ is not fully occupied, and the fast lowering of the $\nu p_{3/2},\nu i_{13/2}$ levels and a rising of the $\nu f_{7/2}$ induced by prolate deformation leads to a large level density around spherical shapes. Hence, large pairing correlations are expected there and, correspondingly, large NMEs are predicted. On the other hand, for larger prolate deformations ($\beta_{2}\in[0.3,0.5]$) we observe that the density of levels around the Fermi level is smaller for protons and neutrons. In particular, gaps are found in the proton s.p.e. produced by the emptying of some levels coming from the $\pi d_{3/2}, \pi d_{5/2}, \pi g_{7/2}$ and the filling of intruder $\pi h_{11/2}$ orbits. In the neutron part, the reduction of the level density is due to the sharp decrease of some $\nu i_{13/2}$ levels, which start to be occupied at $\beta_{2}\sim0.2$, and the increase of some orbits from $\nu h_{9/2},\nu f_{7/2}$.  

The final results for the NMEs and half-lives calculated with D1S and D1M parametrizations are summarized in the Table~\ref{table2}. We observe smaller values of the NMEs for the D1M parametrization although the difference between the results is not very significant. The largest NME corresponds to the decay of $^{152}$Gd that, together with a large resonance enhancement, results in the shortest half-live among the candidates studied here $(T_{1/2}\sim10^{29}$ y). In the evaluation of the half-lives we assume a neutrino Majorana mass $m_{\beta\beta}=50$ meV and the values of $\langle F_{ab}\rangle$, $\Delta$ and $\Gamma_{ab}$ are taken from Ref.~\cite{arXiv:1111.6862v1} corresponding to the electronic transition that gives the shorter half-life in each case. 
We also compare the NMEs obtained in this work with the most recent deformed QRPA calculations~\cite{arXiv:1111.6862v1}. Our values are a factor $\sim3-4$ smaller than the QRPA ones. However, larger NMEs are obtained within our approach if we neglect the configuration mixing and assume single deformations for initial and final states similar to the ones used in QRPA calculations. This shows again the relevance of the deformation and configuration mixing in determining the value of the NMEs.  
\section{Conclussions}
In summary, we have used a state-of-the-art energy density functional method to compute nuclear matrix elements for the most promising $0\nu$ECEC candidates. Our approach includes particle number and angular momentum symmetry restoration and shape mixing along the axial quadrupole deformation $\beta_{2}$, using Gogny D1S and D1M as the underlying interactions. We observe a strong reduction of the NMEs whenever the deformation of mother and daughter nuclei is not close to the spherical shape. All isotopes studied here have a large prolate deformation, and therefore, we predict very small values of the NMEs. Finally, for all cases considered in this work we find half-lives longer than $10^{29}$ years despite the large resonance enhancement found in very recent experiments for these nuclei. 
\begin{acknowledgments} This work was partly supported by the Helmholtz International Center for
FAIR within the framework of the LOEWE program launched by the State of
Hesse and by the Deutsche Forschungsgemeinschaft through contract SFB~634.
\end{acknowledgments}

\end{document}